\documentclass[aps,prl,twocolumn,groupedaddress]{revtex4-1}

\usepackage{dsfont}
\usepackage{color}

\usepackage{graphicx}
\usepackage{dcolumn}
\usepackage{bm}

\newcommand{\be}{\begin{eqnarray}}
\newcommand{\ee}{\end{eqnarray}}

\newcommand{\ket}[1]{\left| #1 \right\rangle}               

\newcommand{\mean}[1]{\left\langle\,#1\,\right\rangle}

\newcommand{\yb}{$^{171}$Yb$^+$}

\begin{document}

\title{Designer Spin Pseudomolecule Implemented with Trapped Ions in a Magnetic Gradient
 }
\author{A. Khromova}
\author{Ch. Piltz}
\author{B. Scharfenberger}
\author{T. F. Gloger}
\author{M. Johanning}
\author{A. F. Var\'{o}n}
\author{Ch. Wunderlich}
\email{wunderlich@physik.uni-siegen.de}

\affiliation{Department Physik, Naturwissenschaftlich-Technische
Fakult\"at, Universit\"{a}t Siegen, 57068 Siegen, Germany}

\date{11 November 2011}

\begin{abstract}

We report on  the experimental investigation of an individual
pseudomolecule using trapped ions with adjustable magnetically
induced $J$-type coupling between spin states. Resonances of
individual spins are well separated and are addressed with high
fidelity. Quantum gates are carried out using microwave radiation in
the presence of thermal excitation of the pseudomolecule's
vibrations. Demonstrating Controlled-NOT gates between non-nearest
neighbors serves as a proof-of-principle of a quantum bus employing
a spin chain. Combining advantageous features of nuclear magnetic
resonance experiments and trapped ions, respectively, opens up a new
avenue towards scalable quantum information processing.

\begin{description}
\item[PACS numbers]
\verb+03.67.Lx 03.65.Ud 37.10.Ty 42.50.Dv+
\end{description}

\end{abstract}


\maketitle

Successful experiments with molecules using nuclear magnetic
resonance (NMR)
\cite{Cory1997,Chuang1998,Jones1998b,Glaser2000,Knill2000,Vandersypen2001,Suter2008}
and with trapped ions
\cite{Cirac1995,Blatt2008,Sackett2000,Kielpinski2002,Kirchmair2009}
have been an important driving force for quantum information
science. Early on during the development of quantum information
science NMR was successfully applied to carry out sophisticated
quantum logic operations and complete quantum algorithms based on
the so-called $J$-coupling between nuclear spins in molecules. In
molecules used for NMR,  the direct dipole-dipole interaction
between nuclear spins  is usually negligible. However, nuclear spins
interact indirectly via  $J$-coupling which is mediated by bonding
electrons.  This $J$-coupling provides a mechanism to implement
conditional quantum dynamics with nuclear spins that are
characterized by long coherence times and are manipulated using rf
radiation. Scalability of NMR is hampered mainly by the use of
ensembles of molecules making it difficult to prepare pure spin
states. Also, nuclear spin resonances and $J$-coupling between spins
in molecules are given by nature, and thus often are not well suited
for quantum computing.

Here, effective spin-1/2 systems are realized by using long-lived
hyperfine states of trapped atomic ions. The vibrational modes of
this individual ion pseudomolecule \cite{Wineland1987} mediate an
effective $J$-type coupling when exposing trapped ions to a
spatially varying magnetic field  \cite{Wunderlich2002,
Wunderlich2003,McHugh2005}. The constants $J_{ij}$ arising from
magnetic gradient induced coupling (MAGIC) of such an individual
spin-pseudo-molecule can be adjusted through variation of the
trapping potential that determines the frequencies  $\nu_n$ of the
ion crystal's vibrational modes. In addition, the range of
interactions can be tuned by applying local static potentials
\cite{McHugh2005,HWunderlich2009}. Single spins can be addressed in
frequency space, since the magnetic field gradient leads to a
position dependent shift of an ion's resonance frequency
\cite{Mintert2001,Johanning2009}. A further useful feature of
spin-pseudo-molecules is the use of radio-frequency and microwave
radiation for conditional quantum dynamics as opposed to laser light
\cite{Mintert2001,Johanning2009,Ospelkaus2011,Timoney2011}, a
feature that substantially reduces experimental complexity and
contributed to the rapid success of NMR in quantum information
science. In addition, this eliminates spontaneous emission that
otherwise may destroy coherences \cite{Akerman2011}.
Moreover,  $J$-type coupling in a spin-pseudomolecule is tolerant
against thermal excitation of vibrational motion. This substantially
reduces the necessity for cooling trapped ions in order to achieve
high fidelity gates.

Exposing a trapped ion Coulomb crystal  to a spatially varying
magnetic field induces a spin-spin interaction mediated by the
common vibrational motion of the ion crystal
\cite{Wunderlich2002,Wunderlich2003},
\begin{equation}
H = -\frac{\hbar}{2} \sum_{i<j }^{N}J_{ij}\sigma_{z}^{(i)}\sigma_{z}^{(j)},
\label{J}
\end{equation}
where $\sigma_{z}$ is a Pauli matrix and the coupling constants
$J_{ij}=\sum_{n=1}^{N}\nu_n \kappa_{ni}\kappa_{nj}$. This sum
extends over all vibrational modes with angular frequency $\nu_n$,
and $\kappa_{nl}\equiv\frac{\Delta
z\partial_z\omega_{l}}{\nu_n}S_{nl}$ indicates how strongly ion $l$
couples to the vibrational mode $n$, when the spin of ion $l$ is
flipped. Here, $\Delta z=\sqrt{\hbar/2m\nu_n}$ is the extension of
the ground state wave function of vibrational mode $n$, and $\Delta
z \partial_z \omega_{l}=g_F\mu_B b_l/\hbar $ gives the change of the
ion's resonance frequency $\omega_l $ when moving it by $\Delta z$
($b_l$ is the magnetic field gradient at the position of ion $l$,
$\hbar$ is the reduced Planck's constant, $\mu_B$ the Bohr magneton,
and $g_F$ the Land\'{e} factor, {\em e.g.},  $g_F=1$ for
$^{171}Yb^+$ ions in the electronic ground state). The dimensionless
matrix elements $S_{nl}$ give the scaled deviation of ion $l$ from
its equilibrium positions when vibrational mode $n$ is excited. Such
magnetic gradient induced coupling may also be implemented using
electrons confined in a Penning trap \cite{Ciaramicoli2005}. Tunable
spin-spin coupling based on optical dipole forces was proposed in
\cite{Porras2004} and demonstrated in
\cite{Friedenauer2008,Kim2009}.

After outlining how individual spins can be addressed with high
fidelity in what follows, the measurement of the coupling matrix
$\{J_{ij}\}$ for a three-spin-pseudomolecule is described. In
addition, it is shown how coupling constants can be adjusted by
variation of the ion trapping potential. Then, the experimental
realization of controlled-NOT gates between any pair of spins is
described, including a CNOT gate between non-neighboring ions. The
entanglement of spins is proven by measuring the parity (defined
below) of a two-spin state.

Hyperfine levels of the $^{2}S_{1/2}$ ground state of $^{171}Yb^+$
serve as an effective spin-1/2 system \cite{Wunderlich2003}, namely
$\ket{\downarrow_i}$ $\equiv$ $^2S_{1/2}$ (F=0) and
$\ket{\uparrow_i}$ $\equiv$ $^2S_{1/2}$ (F=1, $m_F=+1$), where
$i=$~1, 2, 3 refers to ion $i$. These states are coherently
controlled by microwave radiation near 12.65 GHz in resonance with
the $\ket{\downarrow_i} \leftrightarrow \ket{\uparrow_i}$
transition. For the experiments presented here we load three
$^{171}Yb^+$ ions in a linear Paul trap where the effective harmonic
trapping potential is characterized by the secular radial frequency
$\nu_r=2\pi\times502(2)$ kHz and axial frequency
$\nu_1=2\pi\times123.5(2)$ kHz. Initial preparation in state
$\ket{\downarrow}$ is achieved by optical pumping on the $^2S_{1/2}$
(F=1) $\leftrightarrow$ $^2P_{1/2}$ (F=1) transition near 369 nm,
and state-selective detection is done by registering resonance
fluorescence scattered on the $^2S_{1/2}$ (F=1) $\leftrightarrow$
$^2P_{1/2}$ (F=0) electronic transition. This ionic resonance serves
at the same time for Doppler cooling of the ion crystal. The
population of the center-of-mass (c.m.) mode after Doppler cooling
along the axial direction is $\mean{n_1}\approx150$. Microwave
sideband cooling is applied to attain $\mean{n_1}=23(7)$ (details
will be published elsewhere).

The ions are exposed to a magnetic field gradient along the
$z$~direction that is created by two hollow cylindrical SmCo
permanent magnets plated with nickel and mounted at each end-cap
electrode of the trap with identical poles facing each other. The
total magnetic field amplitude is given by
$B(z)=\sqrt{(B_{0||}+b_\mathrm{pm}z)^2+B_{0\bot}^2}$, where $B_{0||}
= 3.4\times 10^{-4}$~T and $B_{0\bot} = 6.2\times 10^{-5}$~T are
longitudinal and radial components of the bias field at the
coordinate origin defined by the position of the center ion, and
$b_\mathrm{pm}$=$19.0(1)$~T/m is the magnetic field gradient created
by the permanent magnets in the absence of a perpendicular bias
field. The magnetic field gradient $b_l=\partial_zB(z)|_{z=z_l}$
defined at the position $z_l$ of ion $l$ is smaller than
$b_\mathrm{pm}$ and not constant due to the nonzero radial component
$B_{0\bot}$ of the bias field.

The state $\ket{\uparrow}$ is magnetically sensitive and undergoes
an energy shift $\Delta E = g_F \mu_B B$ due to the linear Zeeman
effect, while state $\ket{\downarrow}$ to first order is insensitive
to the magnetic field. Because to the gradient of the magnetic
field, three ions with an inter-ion spacing of  $11.9~\mu$m [Fig.
\ref{fig:Ions_And_Spectrum}(a)] are subject to different energy
shifts resulting in a frequency shift of the resonance
$\ket{\downarrow}\leftrightarrow\ket{\uparrow}$ of approximately
$\Delta f\simeq3$ MHz between adjacent ions [Fig.
\ref{fig:Ions_And_Spectrum}(b)]. This energy shift makes it possible
to address independently the ions in frequency space by using
microwave radiation (or laser light \cite{Wang2009}). The
probability amplitude of exciting a neighboring ion decreases with
the square of the detuning. Here, it is less than $4\times10^{-4}$
(see Supplemental Material \cite{Note1}).

\begin{figure}
\includegraphics[width=\columnwidth]{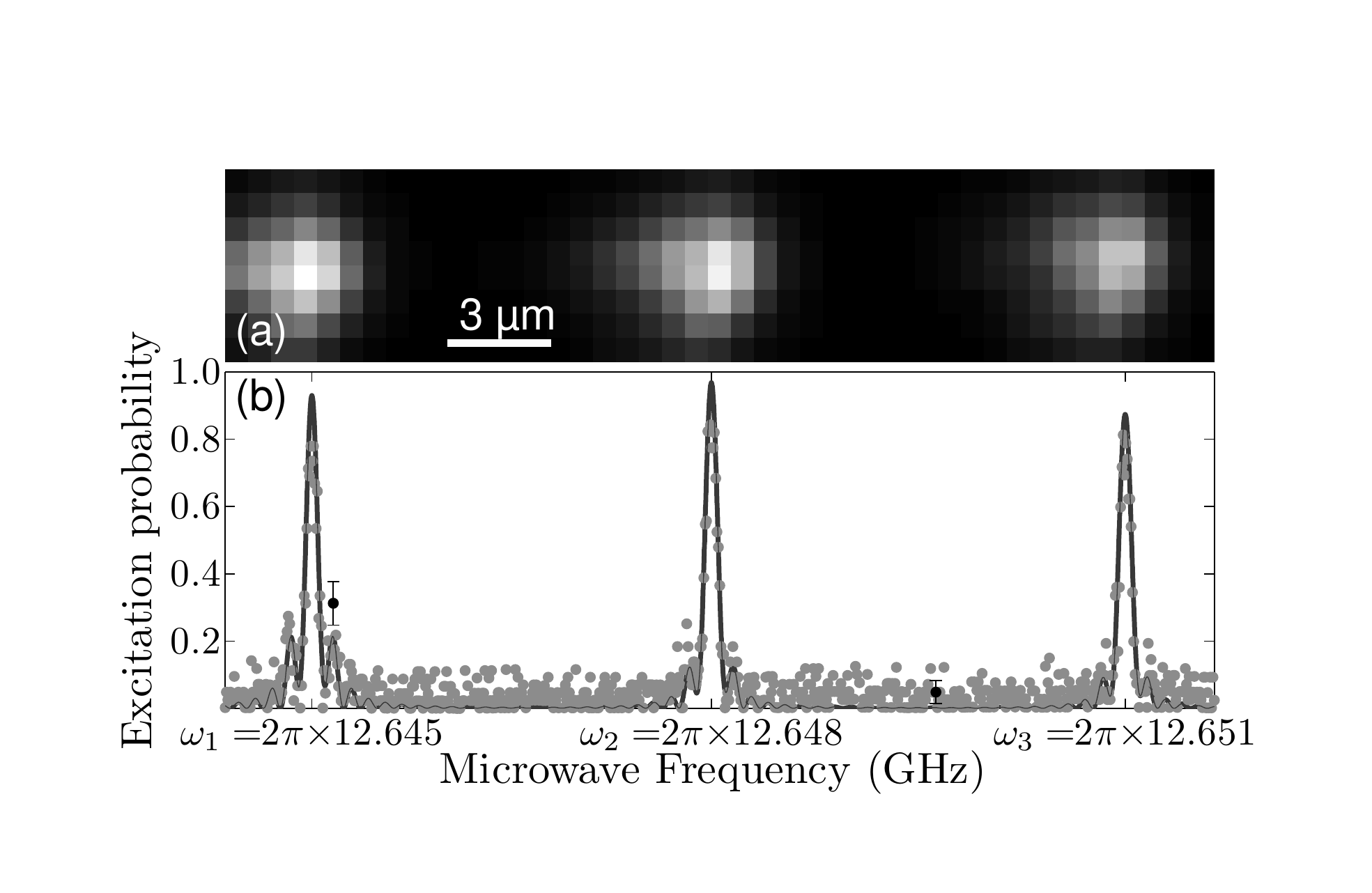}
\caption{ Individual addressing of spins. (a) Spatially resolved
resonance fluorescence (near 369 nm) of three \yb ions recorded with
an intensified CCD camera is shown. Neighboring ions are separated
by 11.9 $\mu$m. (b) Microwave-optical double resonance spectrum of
the above ions. The spectrum was recorded by applying a microwave
frequency pulse of 8 $\mu$s to the ions initially prepared in the
state $\ket{\downarrow}$. The probability to find an ion in
$\ket{\uparrow}$ was determined from counting resonance fluorescence
photons while probing with laser light near 369 nm. In a magnetic
gradient of $\approx18.2$ T/m, the qubit transitions
$\ket{\downarrow} \leftrightarrow \ket{\uparrow}$ of different ions
are nondegenerate.
 The solid line is a fit to the data.
Each data point accounts for
50 repetitions. Two points with error bars are displayed
representing the typical statistical standard deviations.
\label{fig:Ions_And_Spectrum}}
\end{figure}

In addition, the magnetic gradient induces the spin-spin interaction
Eq. (\ref{J}) between the ions' internal states mediated by their
common vibrational modes. Not only nearest neighbors interact but
also the outer ions 1 and 3. The coupling constants $J_{12}$,
$J_{23}$ and $J_{13}$ have been measured in a Ramsey-type experiment
and are displayed in Fig. \ref{fig:Coupling_Matrix}(a) together with
their calculated values. For these measurements, first all three
ions are initialized in state
$\ket{\downarrow\downarrow\downarrow}$. After a microwave $\pi$/2
pulse has been applied to ion $j$, this spin's precession will
depend on the state of  ion $i$ which can be left in state
$\ket{\downarrow_i}$ or set to $\ket{\uparrow_i}$ by a microwave
$\pi$ pulse. After time $\tau$, a second $\pi$/2 pulse with variable
phase $\phi$ is applied and the population $P(\phi)$ of
$\ket{\uparrow_j}$ is measured with ion $i$ initially prepared in
state $\ket{\downarrow_i}$ or in $\ket{\uparrow_i}$, respectively.
The coupling between ions $i$ and $j$ is then deduced from the phase
difference $\Delta \phi_{ij}$ between these two sinusoidal signals
$P(\phi)$: $J_{ij}=\Delta \phi_{ij}/2\tau$. In order to extend the
coherence time of the spin states, which is limited by ambient
magnetic field fluctuations, a multi pulse spin-echo sequence is
applied to ions $i$ and $j$ between the $\pi/2$-Ramsey pulses
\cite{Note1}. The third ion (labeled $k$) has no active role and is
left in state $\ket{\downarrow_k}$ during the whole sequence. Its
interaction with the other ions via $J$-coupling is cancelled by the
applied spin-echo sequence (which is true independent of its
internal state).

It is possible to encode quantum information in two sets of states,
where one set is magnetically sensitive (as is used in this work),
and the other set is not [e.g., $^2S_{1/2}$ (F=1, $m_F=0$) and
($F=0$)]. This allows for temporal storage of quantum information in
magnetically insensitive states that do not couple to other spins
and provide a memory intrinsically robust against ambient field
fluctuations.

\begin{figure}[t]
\includegraphics[width=\columnwidth]{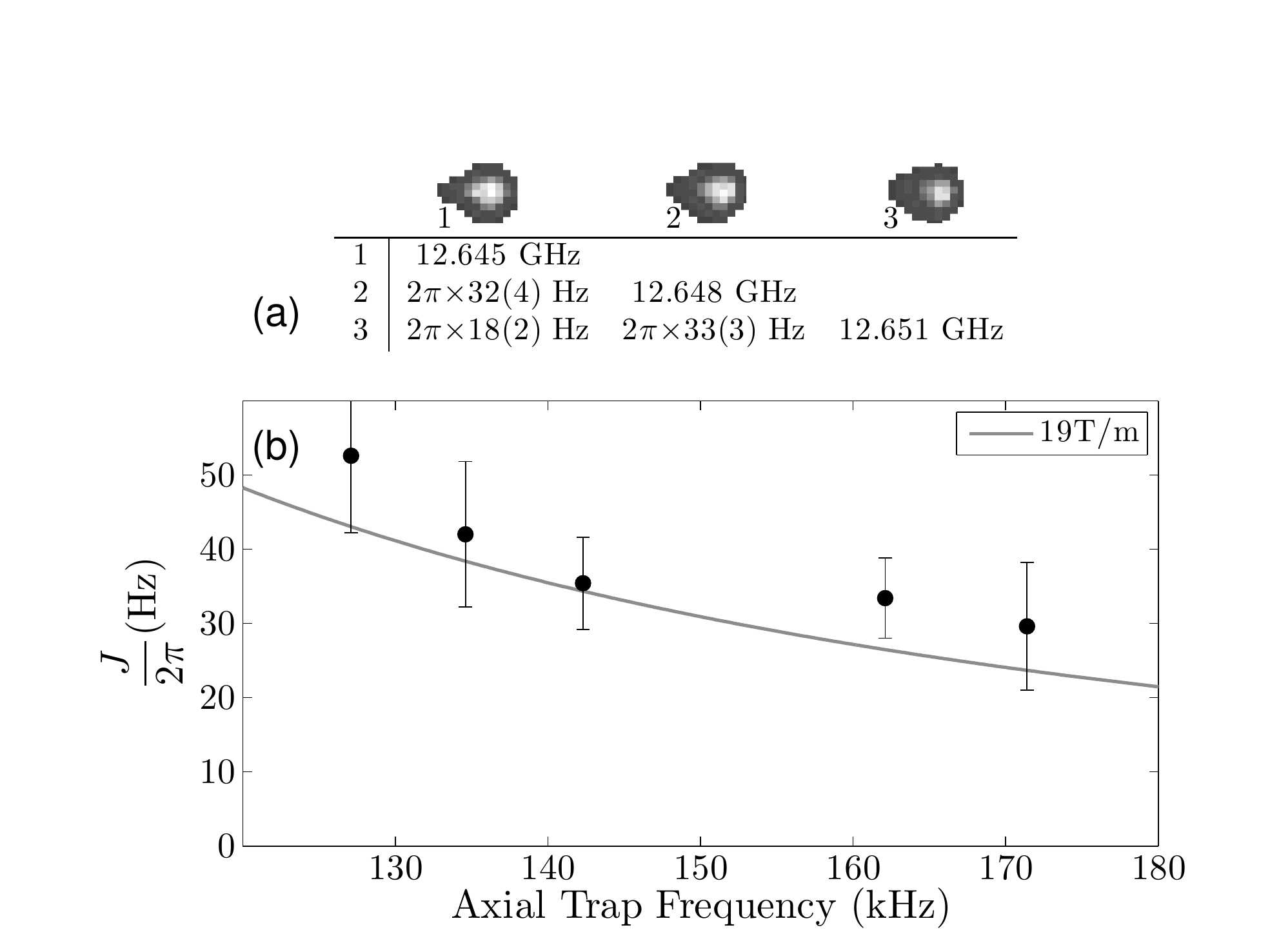}
\caption{$J$-type coupling of a three-spin pseudomolecule. In (a),
the table lists the measured $J$-type coupling constants (below the
diagonal) for a  three-spin pseudomolecule together with the
resonance frequencies of the microwave transitions (on the
diagonal).
For the nonuniform magnetic gradient present in our setup
($b_1=16.8$ T/m, $b_2=18.7$ T/m, $b_3=18.9$ T/m) and the
axial trap frequency ($\nu_1=2\pi\times123.5$ kHz) the calculated
values are $J_{12}=2\pi \times 32.9$ Hz, $J_{23}=2\pi \times 37.0$ Hz, $J_{13}=2\pi \times 23.9$ Hz.\\
(b) Dependence of the coupling strength on the trapping potential.
For a pseudo-molecule consisting of two ions,  the coupling strength
$J$ has been measured for varying c. m. frequency $\nu_1$.  A
calculated curve for a uniform magnetic gradient of 19 T/m is
represented by a solid line. These measurements demonstrate how
$J$-type coupling can be varied by adjusting the trapping potential
\cite{McHugh2005,HWunderlich2009}. \label{fig:Coupling_Matrix}}
\end{figure}

Figure \ref{fig:Coupling_Matrix}(b) shows the dependence of $J$-type
coupling on the c.m. frequency $\nu_1$, that is, on the strength of
the axial trapping potential. These data were taken  with two
trapped ions with the measurements carried out analogous to those
described above, except that only a single spin-echo pulse was used
here. This leads to shorter accessible precession times and thus
smaller phase shifts which in turn yields a larger statistical error
as compared to the data in Fig. \ref{fig:Coupling_Matrix}(a). The
data are in agreement with the calculated dependence of $J$ on
$\nu_1$ ($J\propto (b/\nu_1)^2$).

\begin{figure}[t]
\includegraphics[width=\columnwidth]{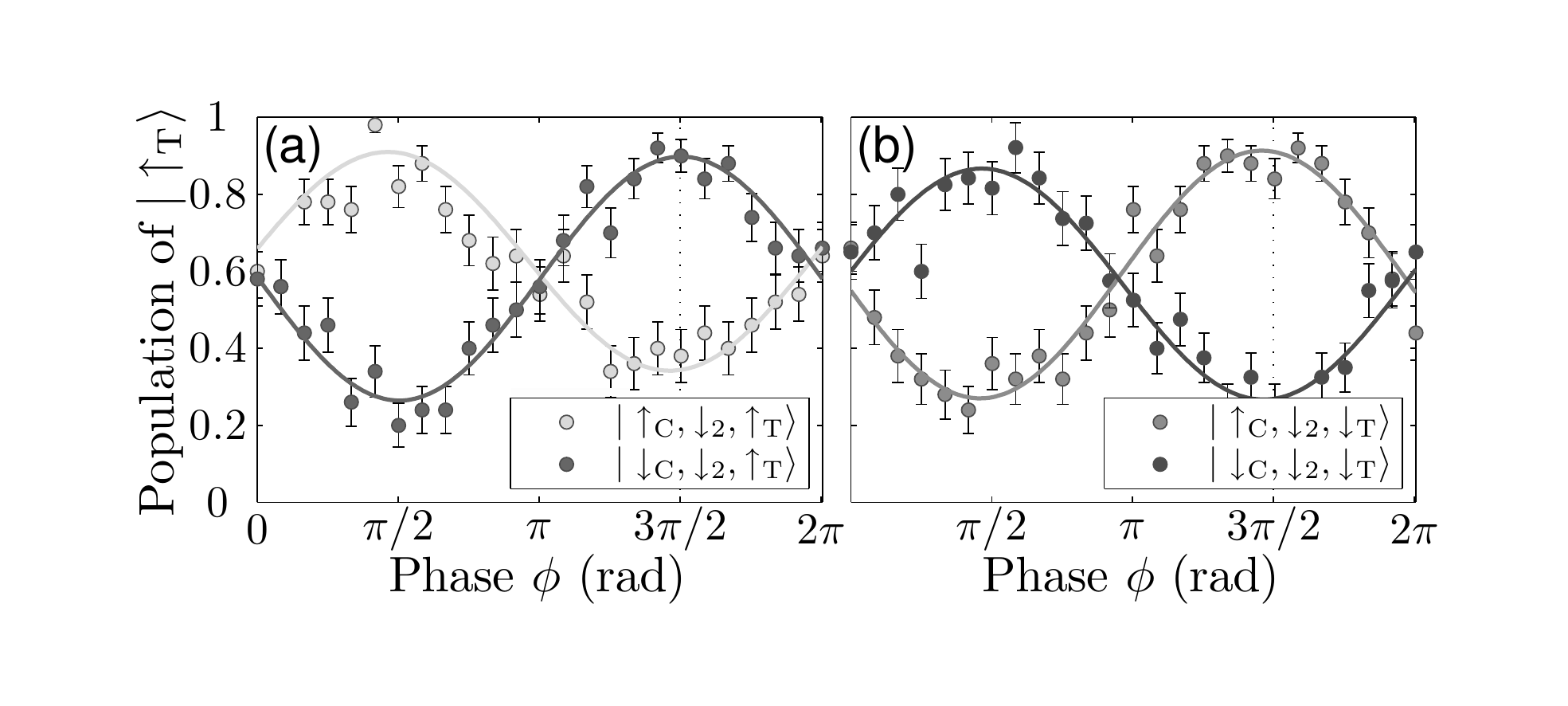}
\caption{Conditional quantum dynamics and CNOT gate between
non-neighboring ions. The probability to find the target spin (ion
3) in state $\ket{\uparrow}$ at the end of a Ramsey-type sequence is
shown as a function of the phase $\phi$ of the second $\pi/2$ pulse
applied to the target. The inset shows the input state prepared
before the Ramsey experiment. In a) the target is initially prepared
in $\ket{\uparrow}$ while in b) it is prepared in
$\ket{\downarrow}$. $J$-type coupling between the control qubit (ion
1) and the target qubit  (ion 3) produces a phase shift on the
target population as a function of the control qubit's state
(different shades of gray). A free evolution time of $\tau=11$ ms
yields a phase shift of approximately $\pi$ ($3.1(2)$ radians)
between the pairwise displayed curves. For $\phi=3\pi/2$, a CNOT
gate results.  The middle ion, prepared in state $\ket{\downarrow}$,
does not interact with other ions. Each data point represents 50
repetitions, the error bars correspond to mean standard deviations
and solid lines are fits to the data. \label{fig:Truth_Table}}
\end{figure}

$J$-type coupling between two spins is employed to implement a CNOT
gate between ion 1 (control qubit) and ion 3 (target qubit). The
evolution time $\tau=11$~ms is chosen to achieve a phase shift of
$\Delta \phi_{13}=\pi$. Figures \ref{fig:Truth_Table}(a) and (b)
show the resulting state population of the target qubit as a
function of phase $\phi$ of the last $\pi/2$-Ramsey pulse which is
applied to the target qubit. The CNOT operation is achieved when
selecting $\phi=3\pi/2$. The four measured sets of data are in
agreement with the truth table of the CNOT gate which induces a flip
of the target qubit or leaves it unchanged depending on the initial
state $\ket{\uparrow}$ or $\ket{\downarrow}$ of the control qubit.

The quantum nature of the conditional gate is verified by creation
of entanglement in the outcome
$\ket{\psi_B}=\frac{1}{\sqrt{2}}(\ket{\downarrow_C\downarrow_2\downarrow_T}
+\mathrm{e}^{i\alpha}\ket{\uparrow_C\downarrow_2\uparrow_T})$ if the
input is a superposition state. Only the correlations of the control
and target qubit determine the parity
$\Pi=P_{\uparrow\uparrow}+P_{\downarrow\downarrow}-(P_{\uparrow\downarrow}+P_{\downarrow\uparrow})$
of the resulting bipartite entangled Bell state ($P_{ij}, \
i,j=\downarrow,\uparrow$ denotes the probability to find the control
and target qubits in the state $\ket{ij}$). When measuring in the
$\sigma_z$ basis, we observe a parity $\Pi_z=0.43(13)$. To prove
that the correlations are nonclassical, the parity $\Pi(\phi)$ was
measured in addition along different bases \cite{Sackett2000}  by
applying additional $\pi/2$ pulses with phase $\phi$ to both ions.
Figure \ref{fig:Parity} shows the resulting signal $\Pi(\phi)$ that
oscillates with twice the phase variation,  as one would expect for
a bipartite entangled state. From the visibility of $V=0.42(6)$ of
the signal shown in Fig. \ref{fig:Parity} we evaluate the fidelity
\cite{Sackett2000} of a Bell state $F=\frac{\Pi_z+1}{4}+\frac{V}{2}$
to be $0.57(4)$ which exceeds the Bell limit of 0.5 and thus proves
the existence of entanglement. This shows that a conditional quantum
gate between two non-neighboring ions is achieved.

\begin{figure}[t]
\includegraphics[width=\columnwidth]{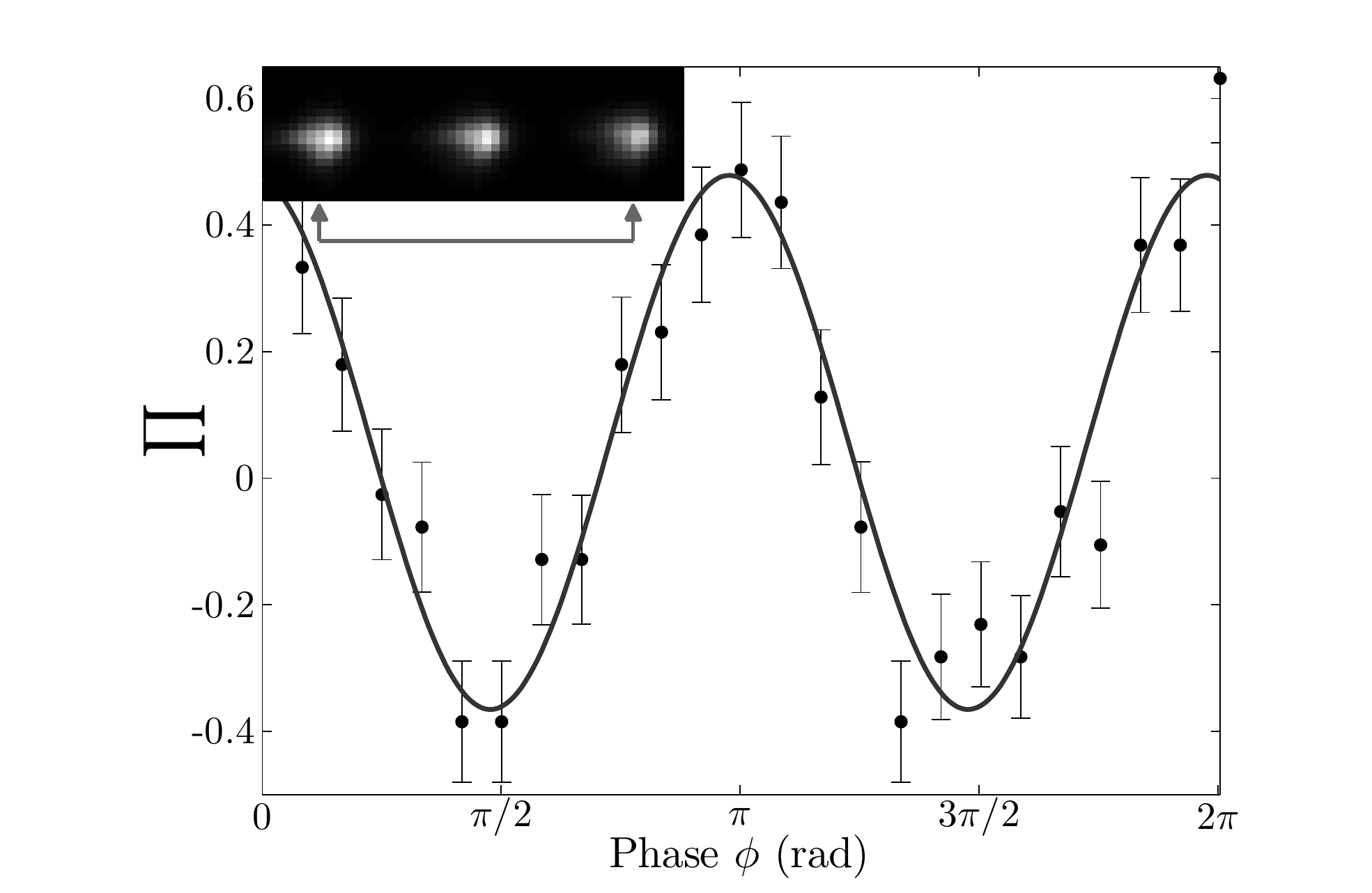}
\caption{Parity signal $\Pi(\phi)$ showing the quantum nature of the
observed correlations of ion 1 and 3. The  Bell state
$\ket{\psi_B}=\frac{1}{\sqrt{2}}(\ket{\downarrow_C\downarrow_2\downarrow_T}
+\mathrm{e}^{i\alpha}\ket{\uparrow_C\downarrow_2\uparrow_T})$ is the
result of a CNOT operation applied to the superposition input state
$\ket{\psi_i}=\frac{1}{\sqrt{2}}(\ket{\downarrow_C\downarrow_2\downarrow_T}
+\ket{\uparrow_C\downarrow_2\downarrow_T})$. In order to measure the
correlations along different bases \cite{Sackett2000}, microwave
$\pi/2$-pulses with phase $\phi$ are applied to both ions followed
by state-selective detection resulting in $\Pi(\phi)$ oscillating as
$\cos(2\phi)$. The fidelity of creating the bipartite entangled
state $\ket{\psi_B}$  is evaluated as $F=0.57(4)$ (see text). Each
data point represents 50 repetitions and error bars indicate 1
standard deviation. \label{fig:Parity}}
\end{figure}

In a similar manner, a CNOT gate was achieved between the first and
the second ion with a fidelity of $F=0.64(5)$ showing that it is
possible to carry out on demand entangling operations between two
ions at desired positions of the ion chain. For neighboring ions the
coupling constants are higher allowing for shorter evolution times
(in this case 8~ms) and therefore reducing the effect of
decoherence. In future experiments, microwave dressed states will be
employed to extend the coherence time of magnetic sensitive states
by several orders of magnitude \cite{Timoney2011}, and thus the
fidelity of quantum gates will be improved.

The entanglement procedure shown here can be applied to longer ion
chains with minimal modifications.  A large-scale quantum processor
would be made up of an array of traps \cite{Kielpinski2002} each
containing a spin-pseudomolecule allowing for simultaneous
conditional quantum dynamics with more than two spins (multiqubit
gates). This could substantially speed up the execution of quantum
algorithms \cite{Schulte2005} and would be an alternative to a
processor that contains zones for conditional quantum dynamics with
two or three ions \cite{Kielpinski2002}. Importantly, physical
relocation ("shuttling") of ions could be avoided during the
processing of quantum information within a given spin-pseudomolecule
\cite{Schulte2005}. The possibility to directly perform logic gates
between distant qubits (e.g., the endpoints of a spin chain as
demonstrated here) makes spin-pseudo-molecules suitable as a quantum
bus connecting different processor regions \cite{Gottesman1999}. In
that case, shuttling would be restricted to a pair of messenger ions
which enable communication between different spin-pseudomolecules.
In addition, a spin-pseudomolecule could serve as a versatile tool
for quantum simulations of otherwise intractable physical systems
\cite{Porras2004,Friedenauer2008,Kim2009,AspuruGuzik2011,Ivanov2011}.

It is desirable to increase the $J$-type constants due to MAGIC. This will be
attained in micro-structured ion traps \cite{Hughes2011,Brueser2011}
and trap
arrays  that allow for the application of larger
magnetic gradients, or by using magnetic field gradients oscillating near
the trap frequency \cite{Welzel2011}. In addition, segmented traps
will allow for shaping $J$-coupling matrices by applying local
electrostatic potentials \cite{McHugh2005,HWunderlich2009}, for example, to
create cluster states \cite{Briegel2009}, or to perform quantum simulations.

Assistance in data-taking by T. Collath and D. Kaufmann and
discussions with  O. G\"uhne are gratefully acknowledged, as well as
financial support by the Bundesministerium f\"{u}r Bildung und
Forschung (FK 01BQ1012), Deutsche Forschungsgemeinschaft, and the
European Commission under STREP PICC and iQIT.


\providecommand{\noopsort}[1]{}\providecommand{\singleletter}[1]{#1}%

\clearpage

\title{Supplement: A Designer Spin-Pseudo-Molecule Implemented with Trapped Ions in a Magnetic Gradient
 }
\author{A. Khromova}
\author{Ch. Piltz}
\author{B. Scharfenberger}
\author{T. F. Gloger}
\author{M. Johanning}
\author{A. F. Var\'{o}n}
\author{Ch. Wunderlich}
\email{wunderlich@physik.uni-siegen.de} \affiliation{Department
Physik, Naturwissenschaftlich-Technische Fakult\"at, Universit\"{a}t
Siegen, 57068 Siegen, Germany}

\date{\today}

\maketitle

\section{Supplemental Material}

In what follows the experimental pulse sequence used for the CNOT
gate is explained. Imperfections that presently limit the fidelity
of conditional quantum gates are also discussed.

{\bf CNOT Gate Pulse Sequence} In order to extend the coherence time
which in this experimental setup is currently limited to about 200
$\mu$s, a sequence of $\pi$-pulses sandwiched between the two
$\pi/2$-Ramsey pulses is applied to both ions participating in the
gate. This serves to refocus dephasing produced by ambient
fluctuating fields. If these $\pi$-pulses were applied only to the
target ion,  they would not only refocus unwanted dephasing, but
also the interaction with the control ion. Therefore, in order to
keep the $J-$type interaction between the ions active (while
compensating for errors caused by ambient fields) the control ion
needs to be flipped each time the target ion is flipped, that is,
the same sequence of $\pi$-pulses has to be applied simultaneously
on both ions.

The sequence of $\pi$-pulses is a novel variant of the
Carr-Purcell-Meiboom-Gill (CPMG) multipulse spin echo method
\cite{Meiboom1958}. Instead of using the original proposal of
identical $\pi$-pulses, their relative phases were varied in order
to obtain a self-correcting sequence that is only sparsely
susceptible to experimental imperfections. Using a sequence with 84
pulses turned out to be a good compromise between suppressing
decoherence and robustness. This sequence has also shown to be more
robust than sequences based on uniform and alternating phases
(details to be published elsewhere).

The sequence of pulses does not refocus magnetic field fluctuations
on a timescale faster than $\tau$. Therefore, for increased free
evolution time a reduction of the contrast of the measured signals
is observed (Fig.~3 of the paper). For a free evolution time of
11~ms, the decoherence accounts for a reduction of the final state
fidelity of at least 30\%. For an evolution time of 8~ms this
reduction accounts for about 20\% error. In future experiments
microwave dressed states will be employed to extend the coherence
time  \cite{Timoney2011b}.

{\bf Addressing Errors} While addressing one ion, it is possible to
unintentionally change the state of a neighboring ion. The
non-resonant excitation probability of neighboring ions is smaller
than $\frac{\Omega^2}{\Omega^2+(g_F\mu_B b d/\hbar)^2}$ where
$\Omega$ is the Rabi frequency, $b$ the magnetic gradient and $d$
the separation of two neighboring ions. In the case of an harmonic
trap with axial confinement frequency $\nu_1$ and $N$ ions of mass
$m$ forming a linear chain, the biggest addressing errors arise in
the middle of the chain where the ions are closest together. The
minimum inter-ion spacing is given by
$d_{min}(\nu_1,N)=\sqrt[3]{\frac{e^2}{4\pi\epsilon_0m\nu_z^2}}\frac{2.018}{N^{0.559}}$,
where $e$ is the electron charge and $\epsilon_0$ is the
permittivity of free space \cite{James1998}. In the experiments
reported here, typically $\Omega\approx2\pi\times60$~kHz,
$d\approx12$~$\mu m$, and $b\approx18$~T/m resulting in a spurious
probability of less than $4\times10^{-4}$.

{\bf State Detection} A photomultiplier is used to measure the
parity signal shown in Fig.~4 of the paper. The number of photons
detected is characterised by three Poissonian distributions that
correspond to the number of ions (between zero and two) to be found
in state $\left|\uparrow\right\rangle$.

In 98.5 \% of the cases the correct state is detected when both ions
are in $\left|\downarrow\right\rangle$ (no fluorescence ideally),
88.9\% of the cases are detected correctly when one ion is in
$\left|\uparrow\right\rangle$ (one ion fluoresces), and 85.2\% of
the cases are detected correctly when both ions are in
$\left|\uparrow\right\rangle$ (both ions fluoresce). These detection
probabilities are independent of the number of repetitions and
reduce the visibility of the parity by approximately 14\%. In future
experiments, the photo detection will be improved by measuring the
final state using a fast, spatially resolving detector.

\end{document}